\documentclass[preprint,showpacs,aps,prc,groupedaddress,floatfix]{revtex4}
\usepackage{epsfig}
\usepackage{amsmath,amssymb}
\usepackage{bm}

\newcommand{\hlf}{\mbox{$\frac{1}{2}$}}

\newcommand{\phm}{\phantom{0}}

\newcommand{\beq}{\begin{equation}}
\newcommand{\eeq}{\end{equation}}

\def\nuc#1#2{\relax\ifmmode{}^{#1}{\protect\text{#2}}\else${}^{#1}$#2\fi}

\begin{document}
\graphicspath{{figures/}}

\title{The relationship between undularity and $l$ dependence of the proton optical model potential}
\author{R. S. Mackintosh}
\email{raymond.mackintosh@open.ac.uk}
\affiliation{School of Physical Sciences, The Open University, Milton Keynes, MK7 6AA, UK}

\date{\today}

\begin{abstract} The contribution of collective or reaction channels to a local optical model potential, OMP, can be readily calculated as a dynamical polarization potential, DPP. The resulting local DPPs commonly have undulatory (`wavy') features, often including local regions of emissivity in the imaginary component. We show here that this undularity arises from $l$-dependence of the underlying formal non-local and $l$-dependent DPP. The $l$-independent proton OMPs, that have the same $S$-matrix $S_{lj}$ as phenomenological $l$-dependent potentials, exhibit undulations that are qualitatively similar to undulations of  local DPPs generated by channel coupling. The $l$-dependent phenomenological potentials studied are the potentials that give the best existing fits to the relevant elastic scattering data and the undulatory potentials presented here,
being $S$-matrix equivalent   (i.e.\ having the same $S$-matrix, $S_{lj}$) give exactly the same scattering. In addition, we present calculations strongly suggesting that undularity (`waviness')  is a generic property of $l$-independent potentials that are $S$-matrix equivalent  to $l$-dependent potentials.  Implications for the validity of folding models based on a local density model are noted.

\end{abstract}


\maketitle


\newpage
\section{INTRODUCTION}\label{intro}

Following the proposal~\cite{cordero}  that the nucleon optical model potential (OMP) is explicitly $l$-dependent, it was often suggested that an $l$-independent potential with a suitably modified  radial form would fit the data equally well.
That turns out to be true, but the undulatory (wavy) radial forms that are required to fit precise and wide angular range data
were not anticipated. Three lines of investigation converge regarding the radial form of the nucleon OMP: (i) experimental elastic scattering observables  that could not be precisely fitted with smooth Woods-Saxon (WS), or similar, forms can be fitted precisely with potentials exhibiting undulations,  (ii) the local and $l$-independent dynamic polarisation potentials (DPPs) generated by coupling to reaction or inelastic channels are significantly undulatory, often with emissive regions, (iii)  an $l$-dependent potential model exists that gives fits over a wide energy range to data that cannot be fitted with WS-like potentials.  
References supporting points (i), ii) and (iii) are given in what follows. Arguments that it is a worthwhile enterprise to get precise fits to highly accurate, wide angular range elastic scattering data are presented in Ref.~\cite{epja}. Arguments that the possibility of angular momentum dependence should be kept open, and a general review of angular momentum dependence of nuclear potentials, will be found in Ref.~\cite{1302}.

Which of the alternative forms of potential,  undulatory or $l$-dependent,  is more natural? This is relevant to the application of OMPs in reaction analyses since two $S$-matrix equivalent potentials will generally not lead to the same radial wave function in the nuclear interior. In this work, `$S$-matrix equivalent potentials' are potentials with the same $S$-matrix $S_{lj}$ where  $l$ and $j$  are the partial-wave orbital and total  angular momenta for spin \hlf\  projectiles. Such potentials clearly give exactly the same observables, so fits to elastic scattering alone can not lead to a preference for $l$-dependent or undulatory potentials. Here we attempt to establish a correspondence between $l$-dependence and undularity. The key point is that for any $l$-dependent potential, we can determine an S-matrix equivalent $l$-independent potential.

Undularity not only occurs in precision fits to high quality elastic scattering data:  the local and $l$-independent representations of the dynamical polarisation potentials, DPPs, generated by coupling to transfer or inelastic channels  are generally undulatory. Such DPPs commonly have radial regions where the imaginary part is emissive. For nucleon scattering examples, see e.g.\ Refs.~\cite{mk76,km83,mk85,mk90,nr2018}. Although  emissive regions in DPPs may appear surprising, the nature of their origin ensures that the unitarity limit $|S_{lj}|  \leq 1$ is not broken. 

It is natural to ask  how the undularity of DPPs can be interpreted, and how it can be  linked to fits of scattering data. Regarding the link to data, we  will show that a phenomenological $l$-dependent potential model, successfully applied to the elastic scattering of 30 MeV protons from \nuc{16}{O}, \nuc{40}{Ca}, \nuc{58}{Ni} and \nuc{208}{Pb}, has undulatory $l$-independent equivalents at each energy, the  imaginary terms often having emissive regions. These potentials have features that are similar to features  appearing in local DPPs resulting from inelastic-channel or reaction-channel coupling, Refs.~\cite{mk76,km83,mk85,mk90,nr2018}.  

In this work we apply  $S$-matrix inversion to determine the $l$-independent potentials that are $S$-matrix equivalent to $l$-dependent proton potentials that fit elastic scattering data. The resulting $S$-matrix equivalent $l$-independent  potentials have qualitative features, namely undularity  and regions of emissivity, that are also found in local and 
$l$-independent DPPs  arising from channel coupling. Such $l$-independent DPPs  are thus the  equivalents of $l$-dependent potentials. This is reasonable since  $l$-dependence is a property of the formal DPPs, as in the theory of Feshbach~\cite{feshbach},  see also Rawitscher~\cite{rawit87}.  As mentioned above, potentials with undulatory features are found when fitting light-ion elastic scattering data that is both precise and having a wide angular range, see Section~\ref{fits}. The occurrence of undulations, and the lack of any widely understood interpretation of them, may inhibit the exact fitting of such data.  Data of that quality arguably contain information concerning the dynamics of nucleon-nucleus and nucleus-nucleus interactions~\cite{epja}.

Every $l$-dependent potential has an $l$-independent equivalent which can be found using $S_{lj} \rightarrow V(r) + {\bf l \cdot s}\, V_{\rm SO}(r) $ inversion,  see Section~\ref{inversion}. Section~\ref{study} presents systematic properties of the $l$-independent potentials that have the same $S_{lj}$ as $l$-dependent potentials precisely fitting proton scattering. Section~\ref{query} discusses issues arising. Section~\ref{inverted} presents the radial properties of the inverted potentials.  The $l$-dependent potentials in earlier sections give precise fits to elastic scattering data, but in Section~\ref{generic}   $S$-matrices for potentials having simpler  forms of $l$ dependence are inverted in order to assess whether undularity, including emissivity, is  a generic property of potentials that are $S$-matrix equivalent to $l$-dependent potentials. Section~\ref{fits} discusses direct model-independent fits to elastic scattering. Section~\ref{mass} relates $l$-dependence to nuclear size. Section~\ref{conc} is a summary and Section~\ref{appendix} is an appendix specifying the characteristics of the $l$-dependent potentials. 

\section{$S$-matrix Inversion}\label{inversion}
The $S$-matrices are inverted using the iterative-perturbative, IP,  $S_{lj} \rightarrow V(r) + {\bf l \cdot s}\, V_{\rm SO}(r) $ inversion  algorithm which is presented in Refs.~\cite{MK82,CM89,Kuk04,pedia}. The IP inversion is implemented in  the inversion code IMAGO~\cite{imago} which quantifies the difference between the $S_{lj}^t$ to be inverted and the $S_{lj}^i$ of the inverted potential in terms of  the S-matrix distance $\sigma$ defined as
\begin{equation} \sigma^2 = \sum_{lj}  | S_{lj}^t - S_{lj}^i|^2. \label{sigma} \end{equation}
The IP iterations  start from a `starting reference potential', SRP, which in all cases presented here was the
$l$-independent part of the $l$-dependent potential. The plots of the inverted potentials presented here are produced by IMAGO and include the SRP. The contribution of the $l$-dependence to the inverted  $l$-independent potential thus appears as the difference between the  inverted potential and the  SRP. It has been established that the IP method can yield inverted potentials that are effectively independent of the SRP and the uniqueness of the inverted potential can be tested  by the use of alternative `inversion bases', see Refs.~\cite{Kuk04,pedia}. The figures produced by IMAGO include the values of  $\sigma$ and in cases where two inverted potentials are shown, that with the lower $\sigma$ is generally adopted.  The tendency for undularity to increase as $\sigma$ becomes very small will be addressed in relation to the significance of the potential undulations. 

\section{$l$-independent equivalent of $l$-dependent phenomenology} \label{study}
The $S$-matrix elements $S_{lj}$, (SMEs), have been inverted for the following $l$-dependent potentials:
30.1 MeV protons on \nuc{16}{O}  and 30.3 MeV protons on \nuc{40}{Ca} from Ref.~\cite{kobmac79}, and for 30.3 MeV protons on \nuc{58}{Ni} and \nuc{208}{Pb}
from Ref.~\cite{kobmac81}. These $l$-dependent potentials fitted the data with a precision exceeding that  achieved with other potentials, and vary more smoothly with energy than the best $l$-independent fits. The  \nuc{16}{O}  and \nuc{40}{Ca} cases are notoriously hard to fit, and cannot be fitted with non-undulatory forms. The inverted potentials having the same $S_{lj}$ as the $l$-dependent potentials obviously  reproduce the data equally well.

The $l$-dependent potentials~\cite{kobmac79,kobmac81} all consist of  standard-form $l$-independent potentials to which $l$-dependent real and imaginary surface peaked terms are added, see the Appendix. We invert $S_{lj}$ for the $l$-dependent potential to obtain the $l$-independent equivalent; subtracting from this the $l$-independent part of
the $l$-dependent potential gives an $l$-independent measure of the $l$-dependent effect. Rather than plot all  the resulting difference potentials, we quantify the $l$-dependence in terms of volume integrals as defined by Satchler~\cite{satchler}.
We present the  differences  between the volume integrals of the $l$-independent 
equivalent potentials and the $l$-independent part of the $l$-dependent potential.
The differences in the volume integrals of the real and imaginary central terms  are $\Delta J_{\rm R}$, $\Delta J_{\rm IM}$ with similar notation for the spin-orbit (SO) terms. These are presented in Table~\ref{properties}   for three cases:
(i)  when only the imaginary  $l$-dependent terms is included, (ii)  when only the real $l$-dependent terms is included and, (iii)  when both $l$-dependent terms are included, as required to fit the data. The significance of the additivity of the real and imaginary $l$ dependencies will be addressed in forthcoming work. 

\begin{figure}
\caption{\label{csRe}   For 30.1 MeV protons on \nuc{16}{O},  the solid lines are the angular distribution (above) and analyzing power (below)  with just the real $l$-dependence included. The dashed lines are calculated with the same potential but with the $l$-dependent terms omitted. The associated change in reaction cross-section was very small: just 0.37 mb. }
\begin{center}
\psfig{figure=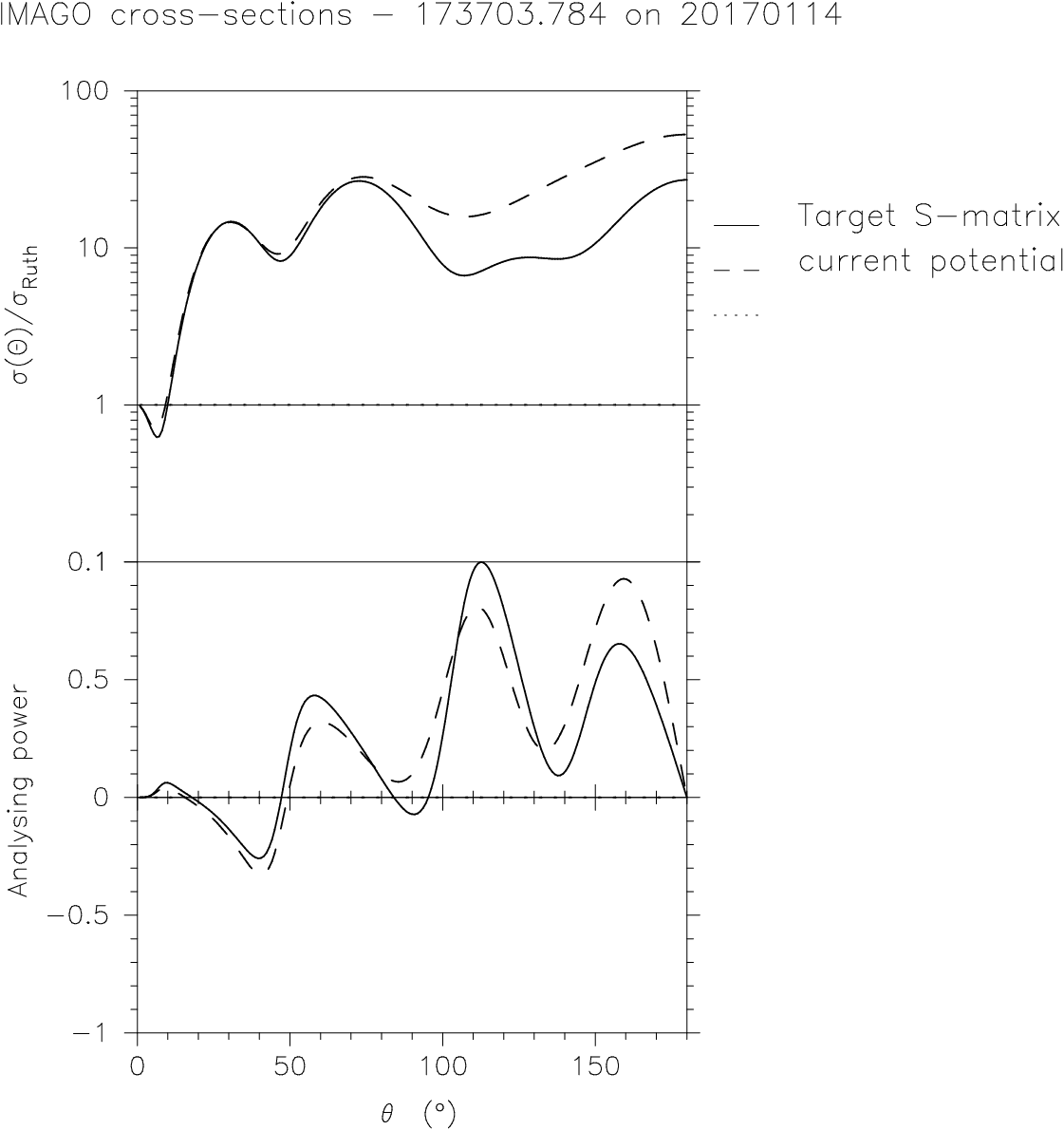,width=10cm,angle=0,clip=}
\end{center}
\end{figure}

\begin{figure}
\caption{\label{csIm}  For 30.1 MeV protons on \nuc{16}{O},  the solid lines are the angular distribution (above) and analyzing power (below) with just the imaginary $l$-dependence included. The dashed lines are calculated with the same potential but with the $l$-dependent terms omitted. The associated change in reaction cross-section was 52.25 mb.}
\begin{center}
\psfig{figure=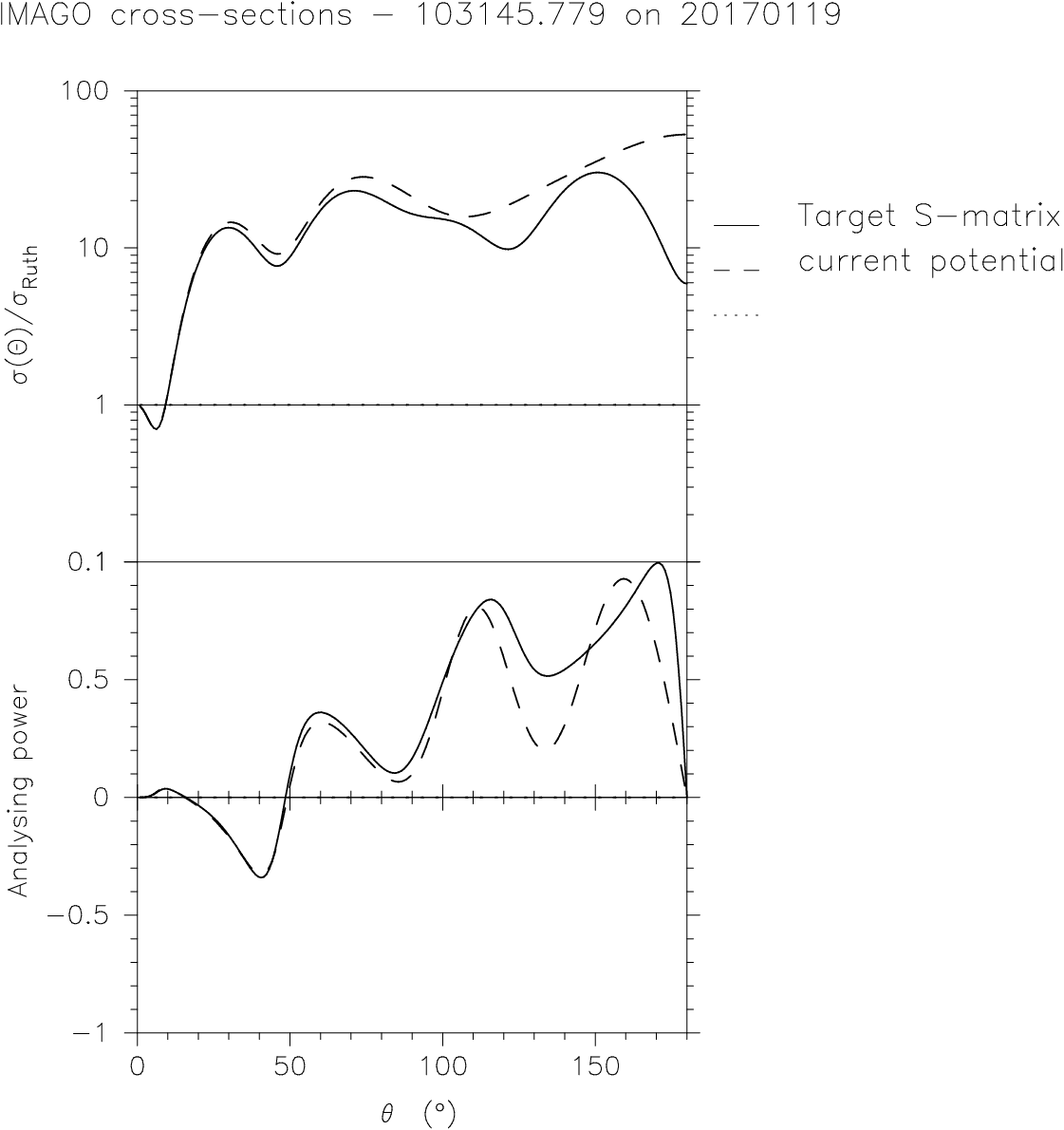,width=10cm,angle=0,clip=}
\end{center}
\end{figure}

\begin{table}[htbp]
\caption{Properties of potentials that are $l$-independent equivalents to $l$-dependent 
phenomenological potentials. For each quantity,  the same quantity for the $l$-independent part of
the $l$-dependent potential has been subtracted, leaving differences $\Delta J_{\rm R}$ and $\Delta J_{\rm I}$ etc.. All volume integrals are in terms of MeV fm$^3$ and the change in reaction cross section, CS, due to the inclusion of the $l$-dependent terms, $\Delta$ CS is in mb. The last column presents the numerical sum of the values in columns 3 and 4.}
\begin{center}
\begin{tabular}{|l|r|r|r|r|r|}
\hline
\phm & Full-Ldep & IM-Ldep & RE-Ldep & $\Sigma$ RE+IM\\ \hline
\multicolumn{5}{|c|}{ p + \nuc{16}{O} 30.1 MeV}\\ \hline
$\Delta J_{\rm R}$	&$-27.25$	&	6.55	&	$-37.21$	&	$-30.66$\\ 	
$\Delta J_{\rm I}$ &	9.4		&11.43	&	0.09		&11.52 \\
$\Delta J_{\rm SOR}$	&0.612&	0.626&	0.289&	0.915\\
$\Delta J_{\rm SOI} $	&0.46&	0.671&	0.011&	0.682\\
$ \Delta$ CS	&55.41	&52.25		&0.37		&52.62\\ \hline

\multicolumn{5}{|c|}{ p + \nuc{40}{Ca}  30.3 MeV}\\ \hline
$\Delta J_{\rm R}$	& 	$-69.21$	&	2.09	&	$-71.59$	&$-69.50$\\
$\Delta J_{\rm I}$     & 	13.649	&	6.439&		7.059	&	13.498\\
$\Delta J_{\rm SOR}$ &		1.4124&		0.0213&		1.4874	&	1.5087\\
$\Delta J_{\rm SOI} $ &	$-0.1451$	&	$-0.0991$	&	$-0.09098$&	$-0.1901$\\ 
$ \Delta$ CS  & 53.13 &		38.81&		20.66&		59.47\\ \hline

\multicolumn{5}{|c|}{ p + \nuc{58}{Ni} 30.3 MeV} \\ \hline
$\Delta J_{\rm R}$	&$-38.85$	&	0.09		&$-40.93$&	$-40.84$ \\
$\Delta J_{\rm I}$     &	8.346	&	0.995	&	6.463	&	7.458 \\
$\Delta J_{\rm SOR}$ &	1.1533	&	$-0.0501$	&	1.3573	&	1.3072 \\
$\Delta J_{\rm SOI} $ &	$-0.1263$	&	$-0.0526$	&	$-0.1578$ 	&$ -0.2104$\\
$ \Delta$ CS  & 29.3	&	6.2	&	23.8	&	30.0 \\ \hline

\multicolumn{5}{|c|}{ p + \nuc{208}{Pb} 30.3 MeV} \\ \hline
$\Delta J_{\rm R}$	&$ -6.82$	&	0.04	&	$-6.59$	 &	$-6.78$\\
$\Delta J_{\rm I}$     &	1.98	&	0.46	&	1.88	 &	2.34 \\
$\Delta J_{\rm SOR}$ &	0.1009	&	$-0.0056$ &		0.1061 	&	0.1005\\
$\Delta J_{\rm SOI} $ &	0.4891	&	0.3829	&	0.0636	&	0.4465\\
$ \Delta$ CS  & 10.1	&	8.3	&	2.3	&	10.6 \\ \hline \hline

\end{tabular}
\end{center}
\label{properties}
\end{table}
The four main sections of Table~\ref{properties}   are headed by an identification of the case, e.g.
30.1 MeV protons on \nuc{16}{O}, etc.
The first column labels what is presented on the line: four sets of changes in volume integrals and one set of changes in the reaction cross section, CS. The remaining four columns present the differences between volume integrals for (i) the S-matrix equivalent $l$-independent potential found for various $l$-dependent potentials, and, (ii) the same volume integrals for the $l$-independent potential to which the $l$-dependent parts had been added. Column 2 presents the effect of the full $l$ dependence, column 3 is for imaginary $l$ dependence
alone, column 4 is for real $l$ dependence alone.  The last column simply adds numbers in columns 3 and 4 for comparison with the corresponding numbers in column 2, reflecting on the linearity of the system to the inclusion of real and/or imaginary $l$-dependent terms.
We note:
\begin{enumerate}
\item For the \nuc{16}{O} case, in column 4, the real $l$ dependence alone resulted (no surprise) in a large change in  $\Delta J_{\rm R}$  and a smaller change in $\Delta J_{\rm I}$. The large change in  $\Delta J_{\rm R}$ is associated with a very small change in the reaction cross section (CS) but simultaneously a very large change in angular distribution (AD) beyond $80^{\circ}$, and in the analysing power (AP) for all angles, see Fig.~\ref{csRe}.

\item Also for \nuc{16}{O}, as expected, the imaginary $l$ dependence gave a somewhat larger volume integral change  $\Delta J_{\rm I}$ than  $\Delta J_{\rm R}$;  for real  $l$ dependence the change $\Delta J_{\rm R}$ was much greater than the change in $\Delta J_{\rm I}$. (In this case, the imaginary $l$-dependent term was at a much larger radius than the real $l$-dependent term~\cite{kobmac79}.) As expected, the change in CS is very large for the imaginary $l$-dependent term, Fig.~\ref{csIm}. The changes in the AD and AP are less than for real $l$-dependence, Fig.~\ref{csRe}, except at angles forward of $80^{\circ}$. This might also be related to the large radius of imaginary $l$-dependent term.

\item The values in the last ($\Sigma$) column are quite close to those in the `Full' column. Exact additivity cannot  be expected.

\item  Points 1 and 2 together exemplify the strong disconnect between the magnitudes of the changes in CS and the changes in the angular observables. This is relevant to evaluating the contribution of fitting the CS when determining OMP parameters.

\item Similar general results apply for  \nuc{40}{Ca}, \nuc{58}{Ni} and \nuc{208}{Pb} targets,  but with diminishing importance of the imaginary $l$-dependence for the heavier targets.  
 
\item In spite of point 5, and the  diminishing effect of $l$-dependence for the heavier targets, the general properties of the $l$-dependence are basically the same for all four target nuclei. This goes back to the claim of Refs.~\cite{kobmac79,kobmac81} that the general properties of the $l$-dependent potential, that `precisely'  fits all the data, vary with energy much more regularly than the best $l$-independent WS-type fits (which fit the data poorly).
\end{enumerate}

In Section~\ref{inverted}, we show that  the  $l$-independent $S$-matrix equivalents to $l$-dependent potentials that actually fit proton elastic scattering have undulations (waviness) in the surface, including regions of emissivity. Since the $l$-dependent potentials are the only potentials that currently fit all the relevant proton elastic scattering data, it follows that the only $l$-independent potentials that currently fit those data have emissive regions in the surface (and undularities not just in the surface.) This is significant since the dynamic polarisation potentials Refs.~\cite{mk76,km83,mk85,mk90,nr2018} arising from various channel couplings exhibit such undularitiles.

\subsection{Higher energy \nuc{40}{Ca}  case}\label{higher}
The $l$-dependent potential in Ref.~\cite{kobmac79} for 35.8 MeV protons on \nuc{40}{Ca} had a somewhat different character from the 30.3 MeV potential since the $l$-independent imaginary part was predominantly of volume character
whereas that for 30.3 MeV had  surface absorption. This case could therefore help answer the question: how does
the character of the inverted $l$-independent potentials depend
on the form of the $l$-independent part of the $l$-dependent potential?   The 35.8 MeV case was studied without spin-orbit terms.

The form of the $l$-dependent term was like that for 30.3 MeV.  Characteristics of the resulting inverted $l$-independent potentials are presented in Table~\ref{properties36}, in the format of  Table~\ref{properties}, for full $l$-dependence as well as for separate real and imaginary $l$-dependence. In Section~\ref{inverted}, the resulting inverted real potential will be seen to have  a form very similar to that for 30.3 MeV,  the imaginary term had very similar undulations (including emissivity) in the surface but with some differences in the nuclear interior. When the real and imaginary terms were added separately, it was found that real $l$ dependence led to a real part that is visually the same  as for full $l$ dependence. However,  real $l$-dependence led to only a very small amplitude undularity in the imaginary potential in the surface, $r>6$ fm, region, although there was some effect for $r<6$ fm.

With $l$-dependence of just the imaginary part, there was little effect on the real potential, but the imaginary potential had a form that was hard to distinguish, except for $r<5$ fm, from that with full $l$ dependence. By eye, the effect of the full $l$ dependence was just the sum of the effects of real and imaginary $l$-dependent components. This is in accord with what is shown in the last column of Table~\ref{properties36}.

\begin{table}[htbp]
\caption{Properties of potentials that are $l$-independent equivalent to $l$-dependent 35.8 MeV
phenomenological potentials for protons on \nuc{40}{Ca}. For each quantity,  the same quantity for the $l$-independent part of
the $l$-dependent potential has been subtracted, leaving differences $\Delta J_{\rm R}$ and $\Delta J_{\rm I}$. All volume integrals are in MeV fm$^3$ and the change in reaction cross section, 
$\Delta$ CS, due to the inclusion of the $l$-dependent terms, is in mb. The last column has the sum of the values in columns 3 and 4.} 
\begin{center}
\begin{tabular}{|l|r|r|r|r|r|}
\hline
\phm & Full-Ldep & IM-Ldep & RE-Ldep & $\Sigma$ RE+IM\\ \hline

\multicolumn{5}{|c|}{ p + \nuc{40}{Ca}  35.8 MeV}\\ \hline
$\Delta J_{\rm R}$	& 	$-69.75$	&1.90 	&	$-72.08$	&$-70.18$\\
$\Delta J_{\rm I}$     & 	13.74	&	9.51&		2.90	&	12.41\\
$ \Delta$ CS  & 10.67 &		8.88&		1.67&		10.55\\ \hline
\end{tabular}
\end{center}
\label{properties36}
\end{table}

\section{Properties and Queries arising}\label{query}
Tables~\ref{properties} and~\ref{properties36} reveal various systematic effects directly related to the properties of the $l$ dependence that was was required to fit the data. The following properrties merit interpretation:\\
{\bf P1} The real $l$-dependence in all cases is such as to reduce the attraction for the lowest partial waves. (Hence negative $\Delta J_{\rm R}$ in all RE-Ldep cases)\\
{\bf P2} The imaginary $l$-dependence in all cases acts to increase the absorption for lowest partial waves (hence positive $\Delta$ CS and positive $\Delta J_{\rm I}$ in all IM-Ldep cases).\\
Note that both these properties are consequences of the $l$-dependent  components of the overall $l$-dependent potentials, and are not related to any comparison with the best (i.e. least worst) $l$-independent potential~\cite{kobmac79,kobmac81} .

Two questions arise:\\
{\bf Q1} Why does $l$-dependence in the \emph{real} part always act to increase $J_{\rm I}$ (positive $\Delta J_{\rm I}$)?\\
Possible answer: from {\bf P1}, the lowest partial waves are subject to less attraction and hence the local wave number is reduced allowing greater absorption along the trajectory. Effectively, the nucleon slows down somewhat and spends more time in the absorptive region. This can easily be seen in terms of the complex momentum of the nucleon within the complex potential. (There will also be refractive effects.) 

{\bf Q2}  Why does the \emph{imaginary} $l$-dependence tend to increase $J_{\rm R}$? \\
Possible answer: As expected from {\bf P2}, the real potential becomes somewhat less effective for small radii where partial waves with low $l$ are most sensitive. But, the effect must not apply for partial waves with large $l$ so the reduced attraction at the centre is compensated by a region of attraction near the surface which  leads to an increase in volume integral. The attractive region near 5 fm connects to undulations in the far surface. Hence the intuition that there should be repulsion is fulfilled for small $r$, but the $r^2$ weighting of the volume integrals wins out, leading to positive $\Delta J_{\rm R}$. 

The tentative nature of the answers suggests that there is much still to be learned about the simplest aspects of nuclear elastic scattering.

\section{Radial Form of the Inverted potentials}\label{inverted}
\begin{figure}
\caption{\label{dppO}   For 30.1 MeV protons on \nuc{16}{O},  the solid lines present the $l$-independent part of the $l$-dependent  potential. The dashed and dotted lines are the inverted potentials, that with dots  having lower inversion $\sigma$ and is the one with properties given in Table~\ref{properties}. 
From top panel downwards, the real-central, imaginary-central, real spin-orbit and imaginary spin-orbit (the last is zero for the $l$-independent term.)  
}
\begin{center}
\psfig{figure=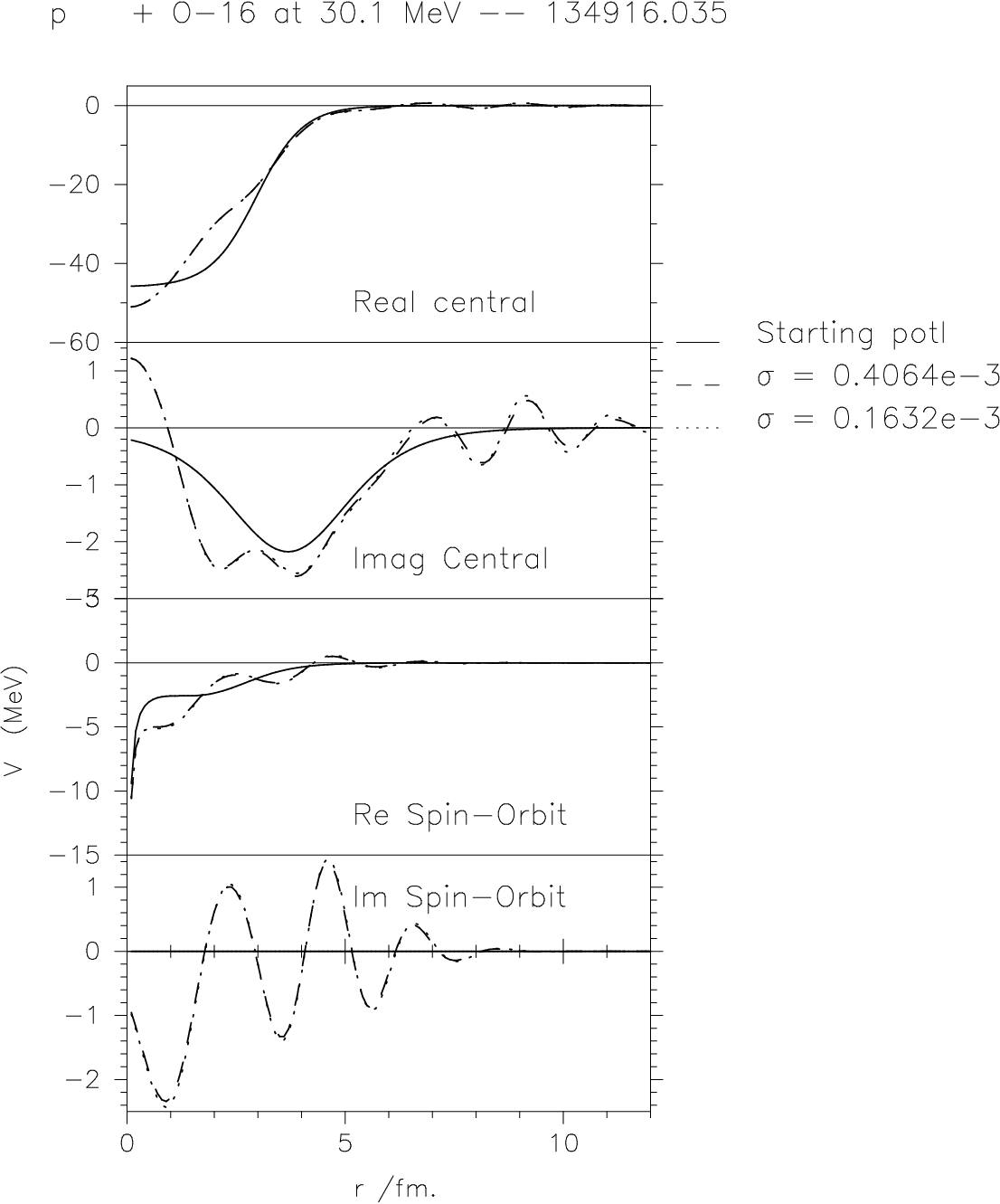,width=11cm,angle=0,clip=}
\end{center}
\end{figure}

\begin{figure}
\caption{\label{dppCa}   For 30.3 MeV protons on \nuc{40}{Ca},  the solid lines present the $l$-independent part of the $l$-dependent  potential. The dashed and dotted lines are the inverted potentials, that with dots  having lower inversion $\sigma$ and is the one with properties given in Table~\ref{properties}. 
From top panel downwards, the real-central, imaginary-central, real spin-orbit and imaginary spin-orbit (the last is zero for the $l$-independent term.
The radial scale is different from that of Fig.~\ref{dppO}.
}
\begin{center}
\psfig{figure=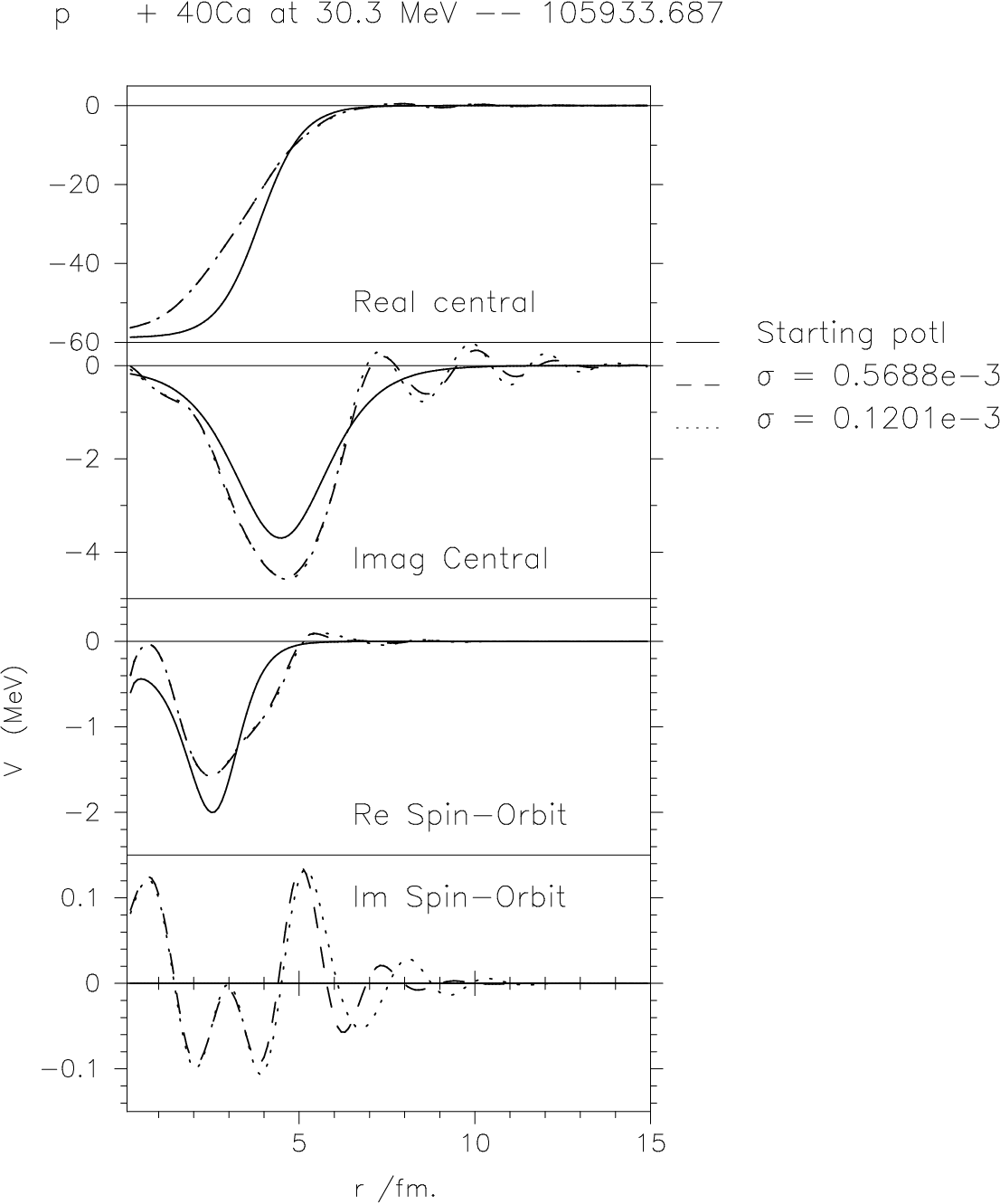,width=11cm,angle=0,clip=}
\end{center}
\end{figure}

\begin{figure}
\caption{\label{dppNi}   For 30.3 MeV protons on \nuc{58}{Ni},  the solid lines present the $l$-independent part of the $l$-dependent  potential. The dotted lines are for the inverted potentials, with properties given in Table~\ref{properties}. The dashed lines follow the solid lines, and the very large initial inversion sigma, 0.926, is given. 
From top panel downwards, the real-central, imaginary-central, real spin-orbit and imaginary spin-orbit (the last is zero for the $l$-independent term.}
\begin{center}
\psfig{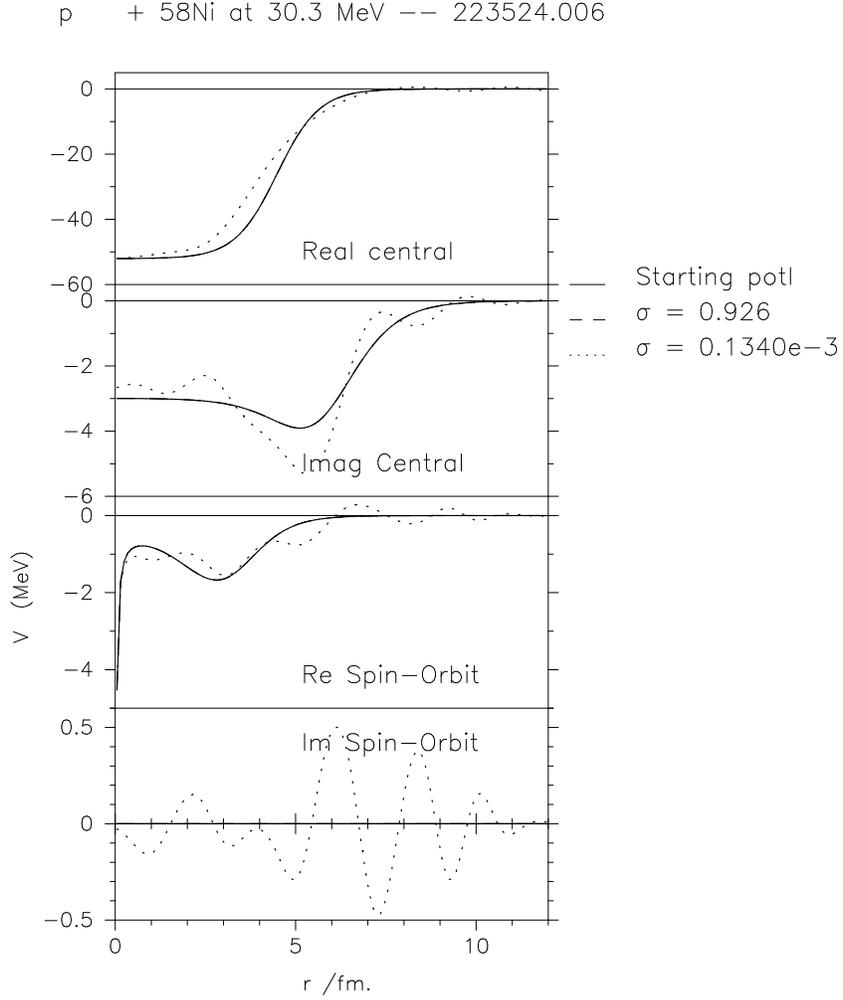}
\end{center}
\end{figure}

\begin{figure}
\caption{\label{dppCa36}   For 35.8 MeV protons on \nuc{40}{Ca}, the solid lines present the $l$-independent part of the $l$-dependent  potential. The dotted lines are for the inverted potential with 
the lowest inversion $\sigma$ and the properties given in Table~\ref{properties36}.  Top panel: the real-central; lower panel: imaginary-central. }
\begin{center}
\psfig{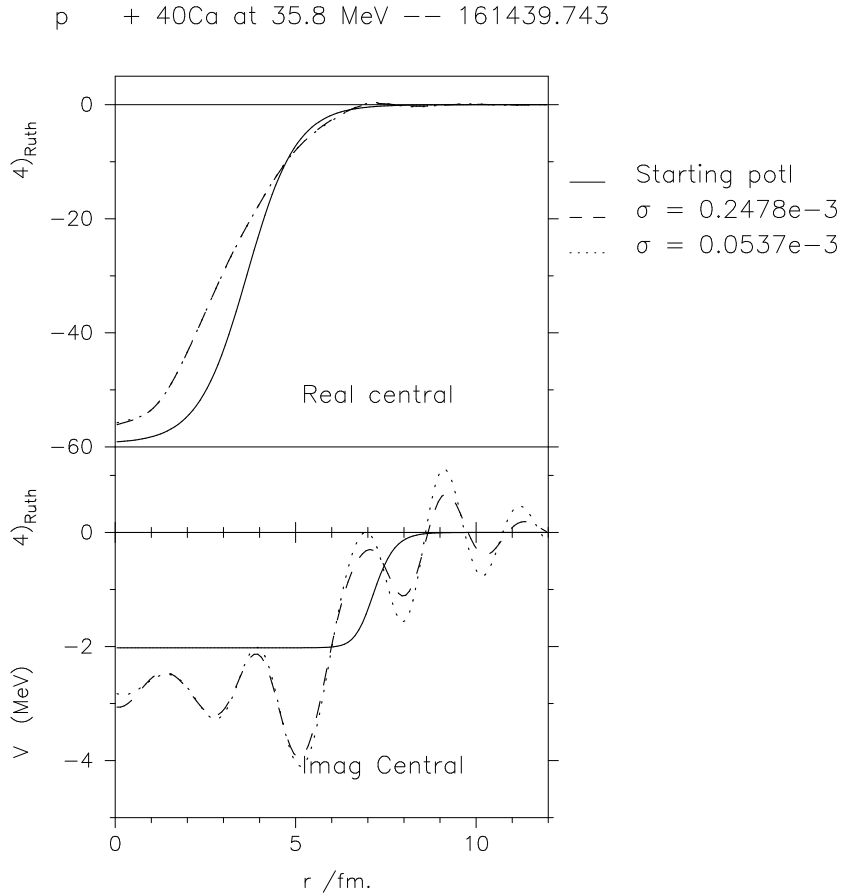}
\end{center}
\end{figure}

Tables~\ref{properties} and~\ref{properties36} present  magnitudes of various properties of the $l$-dependent contribution to the  $l$-independent equivalent potential.  Here we show  explicitly the undularity and emissiveness induced in the $l$-independent equivalent potentials.  The inverted potentials for the \nuc{16}{O}, \nuc{40}{Ca} and  \nuc{58}{Ni} 30 MeV cases are presented in Fig.~\ref{dppO} to Fig.~\ref{dppNi} respectively, and for 35.8 MeV protons on \nuc{40}{Ca} in Fig.~\ref{dppCa36}. In these figures the solid lines represent the $l$-independent part of the $l$-dependent potential and the dashed or dotted lines represent  $l$-independent potentials having the same $S_{lj}$ as the $l$-dependent potential. Thus $\Delta J_{\rm R}$ etc. in Table~\ref{properties} relate to the differences between the solid and dashed or dotted lines.

Two inverted potentials are presented for \nuc{16}{O} and \nuc{40}{Ca}, the lower values of  $\sigma$ corresponding to extended iterations of the IP inversion.  For  \nuc{58}{Ni}  there was just one series of iterations and the dashed line, coincident in this case with the solid line, represents both the $l$-independent part of the $l$-dependent potential and the SRP;  here $\sigma$ has fallen in the inversion process from an initial value of $0.926$  for the SRP  to $0.134 \times 10^{-3}$.

The vertical scales of the various components in Fig.~\ref{dppO} to Fig.~\ref{dppCa36} have been adjusted to the magnitude of each quantity plotted,. Noting this,  the amplitude of the undularity in the far surface of the real central potential is comparable to that of the imaginary potential. Apart  from the \nuc{16}{O} case, the imaginary spin-orbit term (absent from the $l$-dependent potential) is very small, and subject to some uncertainty in the inversion process. The large effect on the real, central term is qualitatively the same for each target nucleus: a strong reduction in depth for smaller radii with a transition to an increase in depth near the surface, leading to undulations further out. The imaginary central potential also exhibits qualitative similarities in all cases, including marked surface undularities which clearly include regions of emissivity. 

For 35.8 MeV protons on \nuc{40}{Ca}, Fig.~\ref{dppCa36} shows that the modifications of the potential due to  the  $l$ dependence are qualitatively the same as for 30.3 MeV case; the change from surface to volume absorption in the $l$-independent part  has made little qualitative difference, especially in the surface region.

Fig.~\ref{dppO} to Fig.~\ref{dppNi}, and also Fig.~\ref{dppCa36}, all exhibit a great reduction in the real potential in the nuclear interior while the change in the surface region is actually somewhat attractive. This leads to an increase of rms radius of the inverted real central: for the 30.3 Mev \nuc{40}{Ca} case   the rms radius of the real part is increased by 0.1568 fm.  In the same case, the rms radius of the imaginary central potential is decreased by 0.2112 fm.  These are quite large changes, but  are intelligible in terms of the properties {\bf P1}  and {\bf P2} above.

\section{Model calculations and generic properties}\label{generic}
It is legitimate to ask whether  undularity is a generic property of  $l$-independent potentials that are $S$-matrix equivalent to potentials that are substantially $l$-dependent. For example, is the occurrence of emissivity in the surface 
a consequence of the particular surface peaked $l$-dependent terms of Refs.~\cite{kobmac79,kobmac81}? It is not easy to give a comprehensive answer to such questions but here the issue is explored with simple model calculations involving an $l$-dependent potential of WS form.   These reveal that a form of $l$ dependence which does not involve a surface peaked $l$-dependent term also leads to $l$-independent equivalents that are strongly undulatory in the surface, including emissive regions in the surface of the imaginary term,  even when only the real them is $l$-dependent. Thus the undulations  in Section~\ref{inverted} are not an artefact of the surface peaked nature of the $l$-dependent component.

There are many ways in which a potential can be $l$-dependent. Here we study just one in which we apply a uniform renormalisation
of the real or imaginary term  for low-$l$ partial waves. The transition has the same dependence on $L$ and $\Delta$ as that for the previous calculations and specified in the Appendix. The real or imaginary part is multiplied by some factor for $l$ values less than $L$ (chosen as specified below), and not modified for high $l$, with a transition region defined by $\Delta$. This is motivated by the possibility that $l$ dependence is characterised by a difference between the interaction for partial waves that have a strong overlap with the nucleus and those that do not. At each energy, the value of $L$ is chosen to be the value of $l$ for which $|S_l|$ is close to $0.5$. Two values of the transition parameter $\Delta$ are chosen to determine whether the amplitudes of the undulations are related to the sharpness of the transition. Here, the spin-orbit terms are omitted. Similar model calculations were carried out, with a similar purpose, in a study~\cite{O16C12} of the angular momentum dependence generated by channel coupling in the case of \nuc{16}{O} scattering from \nuc{12}{C} at 115.9 MeV.

Calculations were performed for 30 MeV and 45 MeV protons scattering from \nuc{40}{Ca}. The Koning-Delaroche~\cite{kd} (KD) potential  was used, without the spin-orbit term. For low partial waves, i.e. for $l$ less than $L$, the real part was reduced by 10 \% and was left unmodified for high $l$. The transition between high and low $l$ was quantified by parameter $\Delta$. Two values, $\Delta = 1$ and $\Delta =2$, were used at 30 MeV to verify that a smaller $\Delta$ leads to stronger undulations. For 30 MeV we set $L= 5.5$ and for 40 MeV we set $L=6.5$. The volume integrals of the real and imaginary potentials 
$J_{\rm R}$ and $J_{\rm I}$ for the inverted $l$-independent potentials are calculated and the corresponding volume integrals for the KD potential are subtracted. The differences,  $\Delta J_{\rm R}$ and $\Delta J_{\rm I}$,  giving a measure of the $l$-independent representation of the effect of the $l$-dependence, are presented in 
Table~\ref{props3045}. The percentage changes are also presented, as are changes in the reaction cross section,  $\Delta$CS.  Consistent with the $l$-dependent modification being confined to the real part, $\Delta$CS is very small although the change in the angular distribution is quite large. These changes are consistent with a uniform decrease in the real phase shift, for $l <L$. There were also changes in $|S_l|$, both positive and negative for different $l$, leading to a small change in CS of about 0.5 mb at 30 MeV and only about 0.2 mb at 45 MeV. However, at 45 MeV, the differential cross section was reduced at all angles except near 150$^{\circ}$ where it was quite small.

These results are presented twice in Table~\ref{props3045} for the 30 MeV cases and three times for the 45 MeV case, corresponding to different values of the $S$-matrix distance  $\sigma$ defined in Eq.~\ref{sigma}. The multiple solutions at each energy bring out a point that is relevant for evaluating the surface undulations.  As the iterations for the $S_l \rightarrow V(r)$ inversion proceed, the undulations in the potential become more pronounced and this is related to the values of $\sigma$  given in Table~\ref{props3045}. The  discussion below refers to potential identifiers, P.I. (pot1 etc.).

\begin{figure}
\caption{\label{45cs3}   For 45 MeV protons on \nuc{40}{Ca},  The solid line represents the angular distribution between 150$^{\circ}$ and 180$^{\circ}$ calculated with the $l$-dependent potential. The dashed line, virtually indistinguishable from the solid line, is calculated with the $l$-independent potential pot3, with inversion $\sigma = 3.79 \times 10^{-5}$. The dotted line corresponds to an earlier iteration, pot1, with $\sigma= 2.30 \times 10^{-4}$.
}
\begin{center}
\psfig{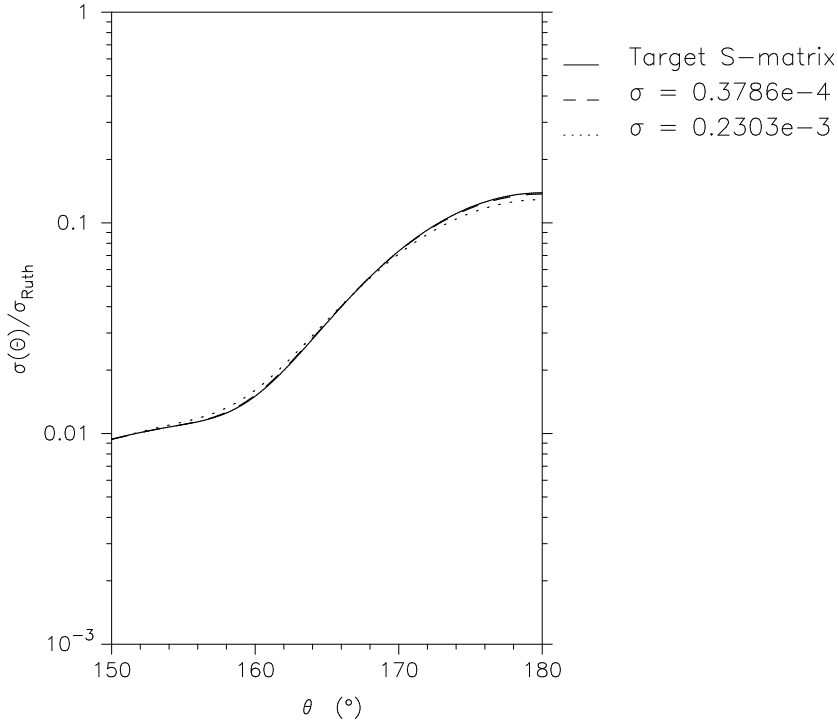}
\end{center}
\end{figure}

The angular distributions for the $l$-independent (inverted) potentials closely fitted that from the $l$-dependent potentials, for 45 MeV, see Fig.~\ref{45cs3}.  The lower value of $\sigma$ corresponds to a small improvement to the fit to the angular distribution, but corresponds to a substantial improvement in the reproduction of $S_l$ for $l$ values that make very little contribution to the angular distribution. Specifically the improvements in fits to $S_l$ were for $l> 11$ at 30 MeV and $l>14$ at 45 MeV. In the latter case, $|S_{14} | \sim 0.9997$ and $\arg S_{14} \sim 10^{-4}$ and the change in the angular distribution corresponding to the correction  of $S_l$  by further iterations was $< 10 \%$ at $180^{\circ}$.

\begin{table}[htbp]
\caption{Properties of potentials that are $l$-independent equivalents to $l$-dependent 
phenomenological potentials. The differences in the real and imaginary volume integrals, $\Delta J_{\rm R}$ and   $\Delta J_{\rm I}$, and the same changes expressed as  percentages, are given.   All volume integrals are in MeV fm$^3$ and the change in reaction cross section, CS, due to the inclusion of the $l$-dependent terms, $ \Delta$ CS is in mb.  The percentage change is also given.  P.I. is the potential identifier.}

\begin{center}
\begin{tabular}{|c|c|c||r|c|r|r|r|r|r|r|}
\hline 
Energy  &$L$ & $\Delta$ &P. I. &$\sigma$&$\Delta J_{\rm R}$ & $\Delta J_{\rm R}$ \% &  $\Delta J_{\rm I}$ 
     &$\Delta J_{\rm I}$ \% &$\Delta$CS  &$\Delta$CS \%  \\  \hline \hline
\multicolumn{11}{|c|}{ Real part $l$-dependent}\\ \hline
     
30.0      & 5.5 & 1& pot2 & $1.66\times 10^{-4}$&$-39.47$ & $-9.56$ &5.09 & 4.44& 0.52 & 0.399 \\ 
30.0      & 5.5 & 1& pot1 & $1.88\times 10^{-5}$ &$-41.12$ & $-9.96$ &10.37 & 9.04& 0.52 & 0.399 \\ \hline
30.0      & 5.5 & 2 & pot1& $4.13\times 10^{-5}$ &$-35.67$ & $-8.64$ &3.41 & 2.97& 0.55 & 0.420 \\
30.0      & 5.5 & 2 & pot2& $3.55\times 10^{-6}$ &$-35.82$ & $-8.69$ &3.92 & 3.42& 0.55 & 0.420 \\ \hline
45.0      & 6.5 & 2 & pot1& $2.30\times 10^{-4}$ &$-32.15$ & $-8.77$ &1.59 & 1.44& 0.207 & 0.208 \\
45.0      & 6.5 & 2 & pot3& $3.79\times 10^{-5}$ &$-32.67$ & $-8.91$ &2.96 & 2.70& 0.206 & 0.207 \\
45.0      & 6.5 & 2 & pot4& $4.02\times 10^{-5}$ &$-32.52$ & $-8.87$ &2.43 & 2.22& 0.206 & 0.207 \\ \hline 
\multicolumn{11}{|c|}{ Imaginary part $l$-dependent}\\ \hline
30.0      & 5.5 & 2 & pot1& $1.41\times 10^{-5}$ &$1.76$ & $0.426$ &8.59 & 7.49& 3.33 & 2.55 \\
30.0      & 5.5 & 2 & pot2& $5.64\times 10^{-6}$ &$1.78$ & $0.431$ &8.58 & 7.48& 3.33 & 2.55 \\ \hline \hline
\end{tabular}
\end{center}
\label{props3045}
\end{table}

\begin{figure}
\caption{\label{45pot3}   For 45 MeV protons on \nuc{40}{Ca},  the solid lines present the  $l$-independent  part of the $l$-dependent potential with the real part in the top panel and the imaginary part below.  The dashed lines represent the inverted potential for the 45 MeV pot1 of Table~\ref{props3045} and the dotted lines  are for the inverted potential pot3 of that table.  The potential pot4 of that table is very close to pot3.
}
\begin{center}
\psfig{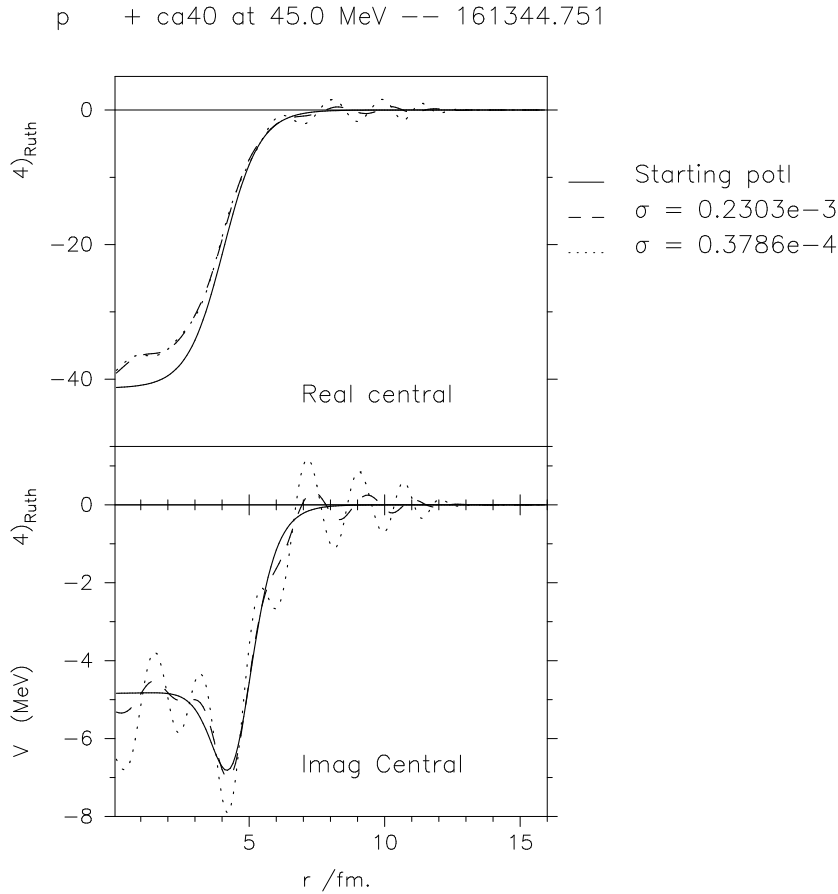}
\end{center}
\end{figure}

The three 45 MeV cases show how improving the fit to $S_l$ for high $l$, thereby lowering $\sigma$, increases the undularity but with only small improvements to the angular distribution. We illustrate this with the pot3 which was the end point of a different sequence of inversion iterations. Fig.~\ref{45cs3} shows that the angular distribution for pot1 differs from that calculated with the $l$-dependent potential by 10 \% at most near 180$^{\circ}$, while the dashed line, representing pot3,  is almost indistinguishable from the solid line.      The potentials pot1 and pot3 are compared with the $l$-independent part of the $l$-dependent potential in Fig.~\ref{45pot3}.

\begin{figure}
\caption{\label{45sm3}   For 45 MeV protons on \nuc{40}{Ca}, the upper panel presents $|S_l|$ and the lower panel $\arg S_l$ for three cases. The solid line is calculated directly from the $l$-dependent potential, the dotted lines are $S_l$ calculated  with $l$-independent potential pot1 and the dashed lines, hard to distinguish from the solid lines, present  
$S_l$ calculated with pot3.}
\begin{center}
\psfig{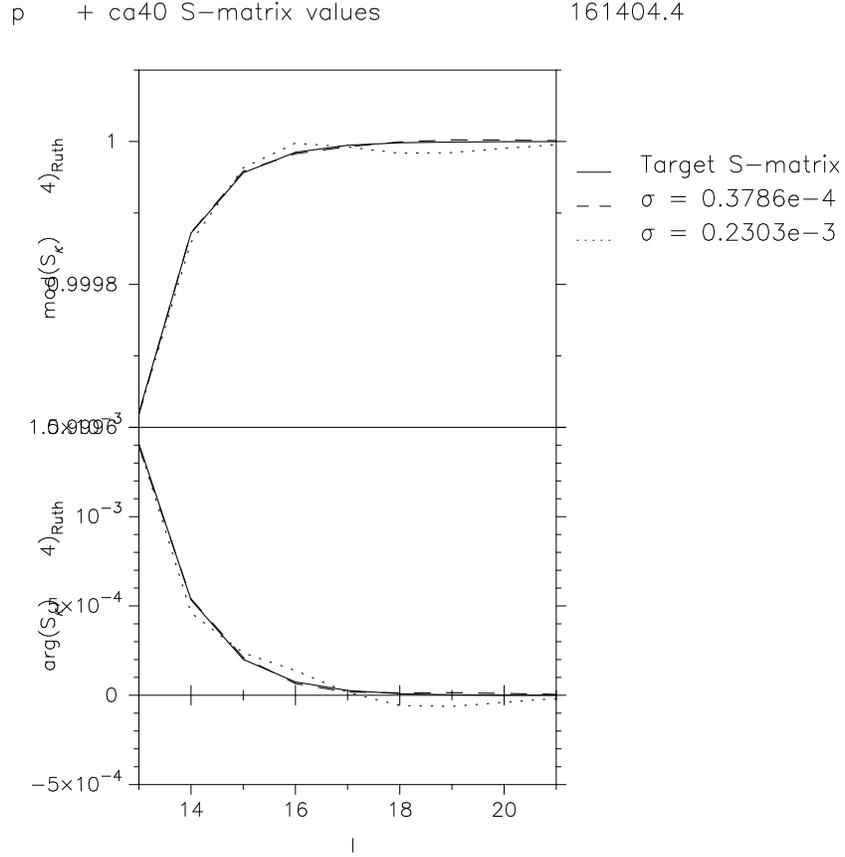}
\end{center}
\end{figure}

The higher value of $\sigma$ for pot1 is due to the poor fit to $S_l$ for high values of $l$, as presented in 
Fig.~\ref{45sm3}.   It is apparent that to reproduce $S_l$ for $l>14$ simultaneously with reproducing $S_l$
for low $l$, larger amplitude undulations are required. Thus an exact representation, with an $l$-independent potential, of  $S_l$ from an $l$-dependent potential of the present form,  requires undulations having an effect on the angular distribution beyond present experimental capabilities.                                                                                                                                                                                                                                                                                                                                                                                                                                                                                                                                                                                                                                                                                                                                                                                           

\subsection{$l$ dependence of the imaginary component}\label{ildep}
To examine the effect of $l$ dependence in the imaginary potential, the imaginary potential for 30.0 MeV is increased by 10 \%  for low partial waves, with $L=5.5$ and $\Delta = 2$.
Inverting was straightforward for this relatively small perturbation yielding potentials having low values of $\sigma$  and characteristics given in the lowest two lines of Table~\ref{props3045}.  

\begin{figure}
\caption{\label{30case4pot1}   For 30 MeV protons on \nuc{40}{Ca},  the solid lines and hidden dashed lines present the  $l$-independent  part of the $l$-dependent potential with the real part in the top panel and the imaginary part below.  The dotted lines represent the inverted potential for the 30 MeV pot1 in the lowest part of of Table~\ref{props3045}.  The potential pot2 of that table is indistinguishable from pot1 out to at least 12 fm.
}
\begin{center}
\psfig{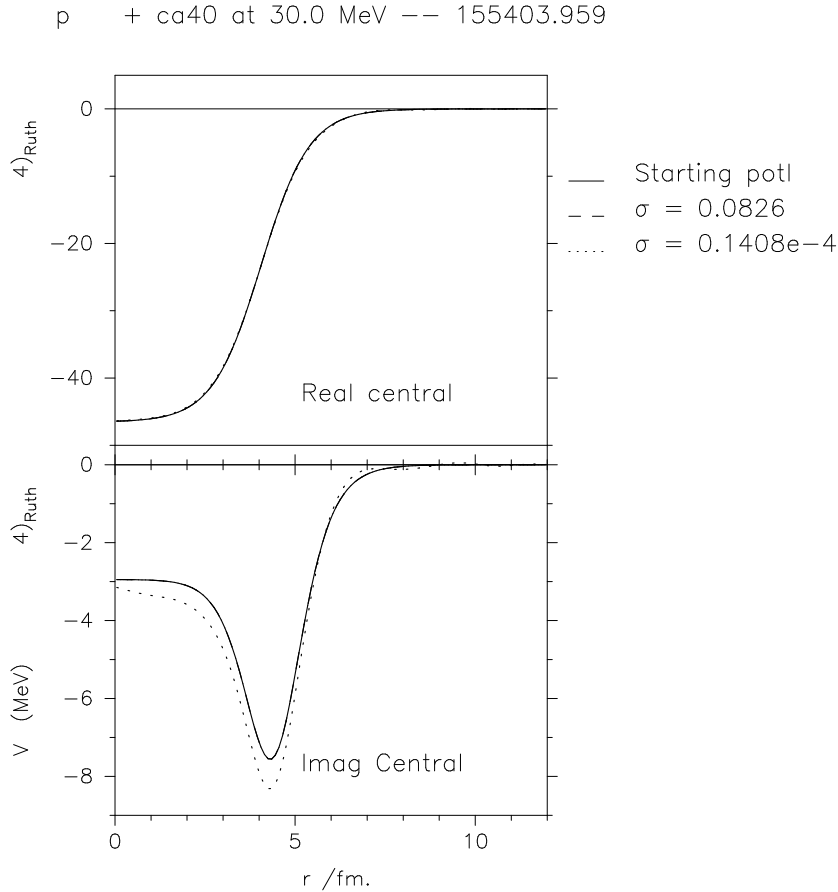}
\end{center}
\end{figure}

\begin{figure}
\caption{\label{30case4pot2}   For 30 MeV protons on \nuc{40}{Ca},  the solid lines represent the  $l$-independent  part of the $l$-dependent potential with the real part in the top panel and the imaginary part below.  The dotted lines and dashed lines represent, respectively, the inverted potentials pot1 and pot2  in the lowest part of Table~\ref{props3045}. The potential pot2 of that table is indistinguishable from pot1 out to at least 12 fm.
}
\begin{center}
\psfig{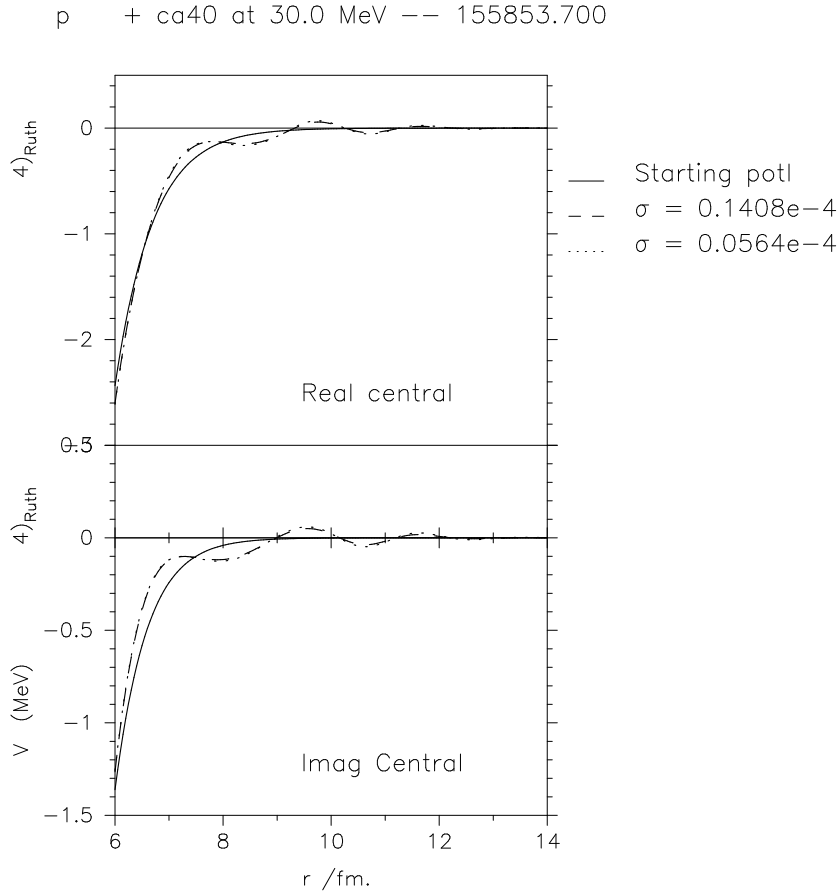}
\end{center}
\end{figure}

Unsurprisingly  the $l$-dependent increase of the imaginary potential led to a much greater increase  in 
the reaction CS  than with the stronger real $l$-dependence. The increase in $J_{\rm I}$ of somewhat less than 10 \% is unsurprising, as is the small percentage change in $J_{\rm R}$.  A uniform 
$l$-independent 10 \% increase
in the imaginary potential would, of course, lead to $J_{\rm I}$ increasing by 10 \% with zero change in  $J_{\rm R}$, and no undulations. There is little change to the real part in the interior region. The undulations resulting from the $l$-dependent increase can be seen in Fig.~\ref{30case4pot1} which also shows that there is an increase in the imaginary potential of roughly 10 \% for $r$ up to about 4.5 fm. For 
larger radii the imaginary potential oscillates about an average of roughly zero change. At around 9 fm, these undulations include an emissive region. This is clearer in the expanded scale of Fig.~\ref{30case4pot2} which also shows that there are undulations in the real part that have a similar amplitude to the undulations in the imaginary part. It is a common feature in these studies to find that surface undulations in either the real or the imaginary part are accompanied by undulations in the other. The potential pot2 improves the fit to $|S_l|$ for between $l=12$ and $l=18$; $|S_l | \simeq 0.99992$ for $l=12$, but makes no visible change to the elastic scattering angular distribution. It appears that the undulations in the surface are driven by the requirement to fit the highest partial waves with a single potential for all partial waves.

\begin{figure}
\caption{\label{30case2cs}   For 30 MeV protons on \nuc{40}{Ca}, the dashed line represents the angular distribution without  any $l$-dependency and the solid line represents the angular distribution for the case of
the real $l$ dependence (10 \% increase) with $L=5.5$ and $\Delta= 2$.}
\begin{center}
\psfig{figure=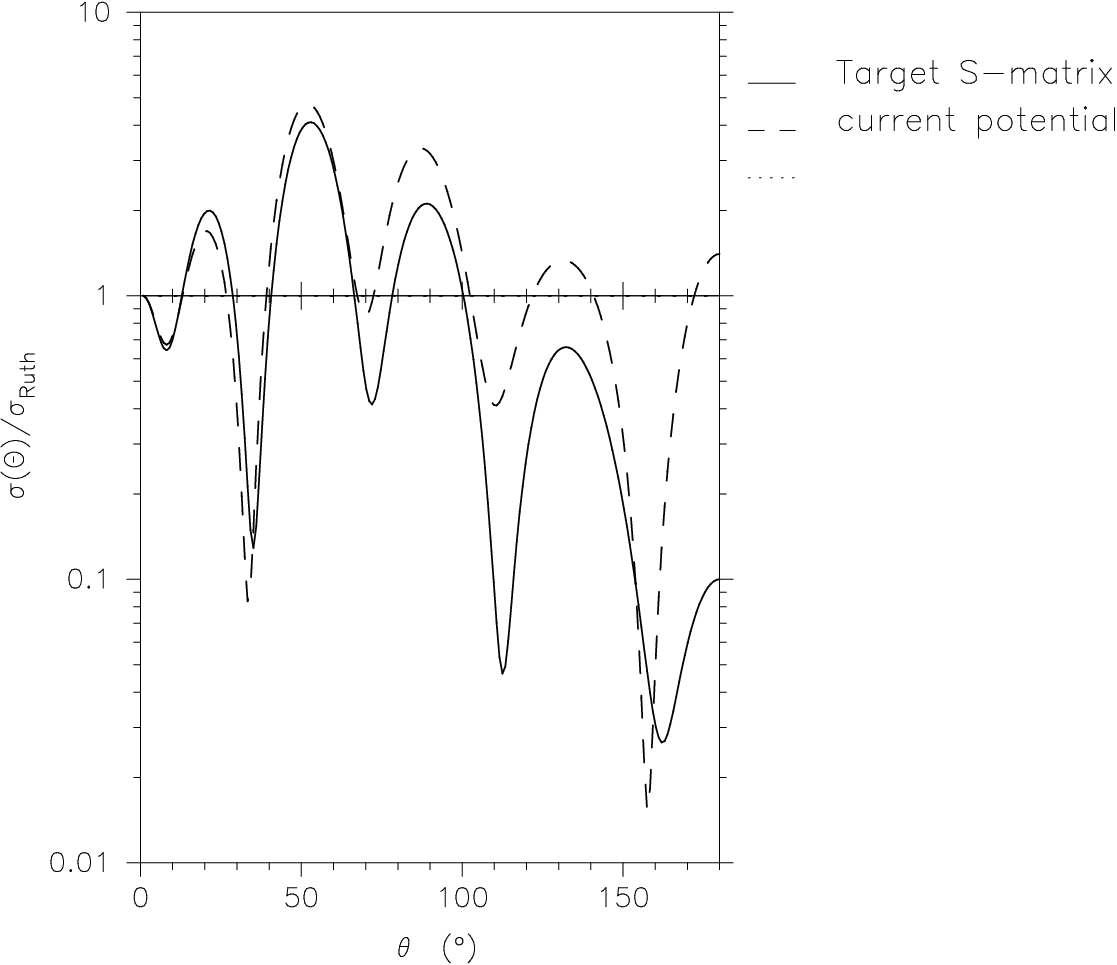,width=11cm,angle=0,clip=}
\end{center}
\end{figure}

\begin{figure}
\caption{\label{30case4cs}   For 30 MeV protons on \nuc{40}{Ca}, the dashed line represents the angular distribution without  any $l$-dependency and the solid line represents the angular distribution for the case of
the imaginary $l$ dependence defined in the text..}
\begin{center}
\psfig{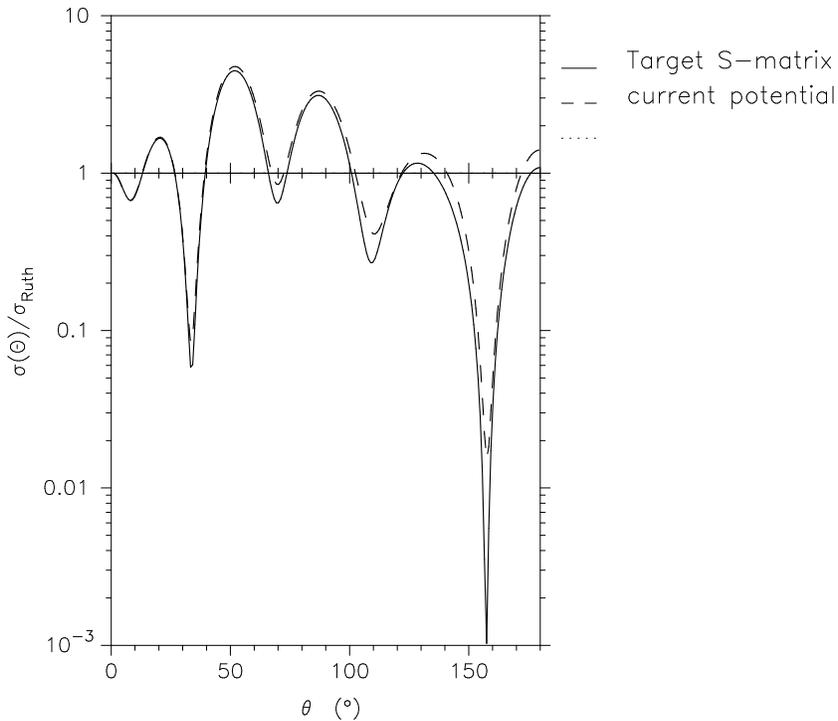}
\end{center}
\end{figure}

The real and imaginary $l$-dependencies have contrasting effects on the reaction cross sections  and the elastic scattering angular distributions. From Table~\ref{props3045} it can be seen that the real $l$-dependent term has quite a small effect on the reaction cross section, CS,  (about 0.42 \% at 30 MeV with $\Delta =2$)  whereas the imaginary $l$ dependence increased CS by 2.5 \%. By contrast, the real $l$ dependence had a large effect on the elastic scattering angular distribution, see Fig.~\ref{30case2cs}, whereas the imaginary $l$ dependence had a much smaller effect on the angular distribution, as in Fig.~\ref{30case4cs}. This is an example of the large disconnect between changes in elastic scattering angular distributions and changes in reaction cross sections. 

\section{$l$-independent fits to elastic scattering data}\label{fits}
Fitting precise, wide angular range, elastic scattering data with an $l$-independent potential leads to a potential having strong undulations. This was shown for protons on \nuc{16}{O} and \nuc{40}{Ca}, see Ref.~\cite{km79}. The fits  to data in this reference were intended to be unprejudiced by theory and were almost model independent. `Almost' model independent because a constraint was imposed to ensure that the imaginary term was not emissive at any radius. This was actually a theoretical prejudice  and the  present work shows that this constraint is a mistake. However,  the general finding that strong undulations are required for an $l$-independent fit stands, although the form of the undulations are distorted by what is now evidently an improper constraint.

Such undulations are not confined to potentials fitting proton elastic scattering.  For deuteron scattering see Refs.~\cite{ermer1,ermer2} and for heavier projectiles see, for example, Refs.~\cite{km82,ksm}. We conclude that, to achieve low $\chi^2$ fits to precise and wide angular range elastic scattering data,  the possibility of undularity must not be excluded. 

The undulations referenced in this section have an amplitude far exceeding variations in the radial density of the target nuclei. The potential undulations are therefore not a direct reflection of undulations in the nuclear density and it is hard to imagine any explanation other than $S$-matrix equivalence to $l$-dependent potentials.

\section{The relationship of $l$-dependence to nuclear mass}\label{mass}
Table~\ref{properties} suggests that the empirically fitted $l$-dependence becomes less important as the nuclear mass increases.
Most calculations of OMPs assume a local density model which does not explicitly take into account  
consequences of the density gradient in the nuclear surface. The $l$ dependence can be seen as a consequence of reaction processes occurring in the presence of gradients in the nuclear density. 
It is then of interest to quantify the comparative importance of the surface region where nuclear density gradients are substantial and examine the relationship between this and the degree of $l$-dependence for nuclei of different masses. We have therefore calculated the ratio $R_{\rm S}$ of the volume integral of the nucleon OMP calculated in two ways:
\beq R_{\rm S} =  \int_0^{\cal R}V(r) r^2 \rm{d}r \,\, [\int_0 ^\infty V(r) r^2 {\rm d}r]^{-1}  \eeq
In the first integral, the upper limit ${\cal R} = (R-a)$ where $R$ is the radius parameter of the Woods-Saxon potential and $a$ is the WS diffusivity. Therefore, $R_{\rm S} $ is a measure of the volume fraction of the OMP that is interior to the surface region, in particular the region where the potential has a substantial radial gradient.

We have calculated  $R_{\rm S} $  for \nuc{16}{O}, \nuc{40}{Ca}, \nuc{58}{Ni} and \nuc{208}{Pb}. In each case, we calculated $R_{\rm S} $   for the real central term of the Koning Delaroche global potential~\cite{kd}. The values  for those four nuclei, were, respectively: 0.254, 0.426, 0.466, 0.642. All of these numbers point to the importance if the surface region, but most particularly \nuc{16}{O} stands out as being in a separate class. If $l$-dependence is, as we propose, related to the influence of surface processes, and strongly influenced by departures from the validity of local density approximation, then we should not be surprised to find strong effects 
associated with $l$-dependence for \nuc{16}{O}. It would be interesting to know what part of the effect due to channel coupling, e.g. to deuteron channels ~\cite{prc98}, is accounted for in local density folding models.

\section{Summary and discussion}\label{conc}
Section~\ref{study} presented characteristic properties of $l$-independent potentials that were found by inversion to be $S$-matrix  equivalents of  $l$-dependent potentials. The particular $l$-dependent potentials give excellent fits to elastic scattering data that cannot be fitted with conventional radial forms.  Section~\ref{inverted} compared the inverted potentials with the $l$-independent part of each $l$-dependent potential and  Figs.~\ref{dppO} to~\ref{dppNi} show that each $l$-independent potential found by inversion has undulations in the surface.  The undulations in the imaginary terms include radial regions where the potential is emissive. This tendency is strongest for light target nuclei and persists to a smaller degree for the \nuc{208}{Pb} case, not shown. Since the $S_{lj}$  for $l$-dependent potentials all  satisfy the unitarity bound, this is true also for the inverted potentials, despite the emissive regions. It follows that $l$-independent potentials that fit elastic scattering of 30 MeV protons from \nuc{16}{O}, \nuc{40}{Ca}, \nuc{58}{Ni} and \nuc{208}{Pb}, probably better than any other potentials, have undularities with emissive regions in the nuclear surface. Since they are the $l$-independent equivalents of $l$-dependent potentials,  it is reasonable to attribute  undularity and emissivity  found in phenomenological potentials  to an underlying $l$ dependence. It is also  reasonable to attribute the appearance of undularity and emissivity in local DPPs from coupled channel calculations, determined by inversion of $S_l$ or $S_{lj}$, to the $l$ dependence of an underlying non-local and $l$-dependent DPP.

Does the strong undularity, involving emissive regions, result from the particular form of $l$ dependence applied in Refs.~\cite{kobmac79,kobmac81}? That particular $l$-dependent term was quite sharply confined to the surface region. The calculations presented in Section~\ref{generic} suggest that the occurrence
of undularity is a generic property  of potentials that are $l$-independent $S$-matrix equivalents of $l$-dependent potentials. In that section the $l$-independent equivalents were found for potentials for which the real or imaginary part was 
multiplied by a factor over a range of partial waves: $l  <  L$ with a transition over a range $\sim \Delta$. For the real part, the factor was 0.9 and for the imaginary part, 1.1. The resulting potentials were undulatory. The undulations in the $l$-independent potential  had regions of emissivity in the imaginary part as a result of $l$ dependence in either the real or imaginary terms. The sharpness of the $l$-dependence transition was adjustable, and a sharper transition (smaller $\Delta$) led to somewhat stronger undulations. A similar relationship between $l$-dependence and undularity was found in the case of \nuc{3}{He} scattering on \nuc{58}{Ni}~\cite{helion} and also found for \nuc{16}{O} scattering from \nuc{12}{C} at 115.9 MeV~\cite{O16C12}.  \emph{It therefore appears that there is a generic  relationship between undulations in optical potentials and an underlying $l$ dependence.}

In Section~\ref{mass} we presented a plausible argument for the $l$-dependence of proton OMPs becoming weaker as the mass of the target nucleus increases, as suggested by the results in Table~\ref{properties}. 

{\bf Implications:} Any necessity for $l$ dependence indicates a failure of a folding model based on a local density approximation. Thus, for any case of elastic scattering, if a precise model-independent, but $l$-independent, empirical fit to wide angular range precise data exhibits undulations,  then the actual potential for that case must be $l$-dependent, indicating a failure of local density folding models. In this situation, it is  unclear what potential is appropriate for use in direct reactions.  This must depend on what channels are included and the use of local $l$-independent potentials is unjustified. Since DPPs never amount to a  uniform renormalization, it is clear also that the appropriate way to evaluate a folding model potential is by fitting a model-independent additive term, when  data of suitable quality exists.

The key point is that imprecise fits to nuclear elastic scattering data of typical range and precision miss key physics; there appear to be historical reasons why precision fitting of nuclear scattering data is not taken as seriously as precision fitting of electron scattering data.  A possible disincentive for treating $l$-dependence is the uncertainty as to how to incorporate the consequences of causality, i.e.\ dispersion relations that must hold between the real and imaginary potentials. This has been studied for a particular form of $l$-dependence in Ref.~\cite{atmr}.

\section{Appendix: specification of the $l$-dependence}\label{appendix}
The $l$ dependence adopted in Refs.~\cite{kobmac79,kobmac81} and of  Section~\ref{study} was essentially that of the original work, Ref.~\cite{cordero}. The $l$-dependent potential was the sum of a standard $l$-independent term with an added $l$-dependent  term. The $l$-independent term was a Woods-Saxon real part plus an imaginary part that was the sum of Woods-Saxon and Woods-Saxon derivative terms. 
A conventional spin-orbit term was included and all terms were as defined by Perey and Perey~\cite{PP76}. 
The $l$ dependency of  was in the form of  an additional central potential having  real and imaginary Wood-Saxon derivative (surface peaked) form with an overall $l$-dependent factor:
\beq f(l) = \frac{1}{1 + \exp{((l^2 -L^2)/ \Delta^2})} . \label{fl} \eeq
Thus, the $l$-dependence has the form of an additional surface potential that cuts off sharply when $l>L$. There is 
the one overall $l$-dependent factor $f(l)$  
 for both real and imaginary parts which had independent radial and strength parameters. There was no $l$-dependence in the spin-orbit term which was real  in almost all cases.   This $l$-dependent potential fitted the data with very consistent parameters, the values of which varied more smoothly with energy than the parameters of the best $l$-independent potentials. In all cases, the searches led to a real $l$-dependent term that was repulsive (i.e. there was less attraction for $l  \ll L$) and an imaginary term that was absorptive (i.e. more absorption for $l \ll L$.) This is immediately reflected in the volume integral results in Table~\ref{properties}.

{\bf Potentials for Section~\ref{generic}.} 
The real $l$-dependence of Section~\ref{generic} was achieved by adding a repulsive real potential of identical form
to the original real potential but of one tenth the magnitude and with overall factor $f(l)$ of Eq.~\ref{fl}. Thus the potential
is reduced by 10 \% for low partial waves. The imaginary $l$ dependence was effected in a similar fashion.

\end{document}